\documentclass[showpacs,doublespace,preprint,amsmath,amssymb,pre]{revtex4-1}

\usepackage{dcolumn}
\usepackage{bm}
\usepackage{graphicx}
\usepackage{color}
\usepackage{hyperref}
\usepackage{latexsym}
\usepackage{amsthm}
\usepackage{verbatim}
\usepackage{setspace}
\usepackage{wasysym}
\usepackage{amsmath}
\usepackage{amsfonts}
\usepackage{amssymb,color}
\usepackage{subfigure}
\usepackage{graphicx,color}
\usepackage{ifpdf}

\begin{document}

\def\be{\begin{equation}}
\def\ee{\end{equation}}

\def\bc{\begin{center}}
\def\ec{\end{center}}
\def\bea{\begin{eqnarray}}
\def\eea{\end{eqnarray}}

\newcommand{\avg}[1]{\langle{#1}\rangle}
\newcommand{\Avg}[1]{\left\langle{#1}\right\rangle}

\newcommand{\cch}[1]{\left[#1\right]}
\newcommand{\chv}[1]{\left \{ #1\right \} }
\newcommand{\prt}[1]{\left(#1\right)}
\newcommand{\aver}[1]{\left\langle #1 \right\rangle}
\newcommand{\abs}[1]{\left| #1 \right|}
\newcommand{\OO}{\cal O}

\def\ie{\textit{i.e.}}
\def\etal{\textit{et al.}}
\def\m{\vec{m}}
\def\G{\mathcal{G}}
\def\fig{Fig.}
\def\tab{Table}

\title{Jamming and percolation of dimers in 
restricted-valence random sequential adsorption}

\author{A. P. Furlan}
\email{apfurlan@fisica.ufmg.br}
\affiliation{Departamento de F\'isica, ICEx,
  Universidade Federal de Minas Gerais, C. P. 702, 30123-970 Belo
  Horizonte, Minas Gerais - Brazil}

\author{Diogo C. dos Santos}
\affiliation{Departamento de Estat\'istica, ICEx,
  Universidade Federal de Minas Gerais, 30123-970 Belo
  Horizonte, Minas Gerais - Brazil}

\author{Robert M. Ziff}
\affiliation{Center for the Study of Complex Systems and 
Department of Chemical Engineering, University of Michigan, 
Ann Arbor, Michigan 48109-2136, USA}

\author{Ronald Dickman}
\affiliation{Departamento de F\'isica and National Institute 
of Science and Technology for Complex Systems, ICEx,
  Universidade Federal de Minas Gerais, C. P. 702, 30123-970 Belo
  Horizonte, Minas Gerais - Brazil}

\date{\today}

\begin{abstract} 
Restricted-valence random sequential adsorption~(RSA) is studied
in its pure and disordered versions, on the square and triangular
lattices. For the simplest case~(pure on the square lattice) we prove 
the absence of percolation for maximum valence $V_{\rm max}=2$. 
In other cases, Monte Carlo simulations are used to investigate the 
percolation threshold, universality class, and jamming limit. Our results
reveal a continuous transition for the majority of the cases studied. The 
percolation threshold is computed through finite-size scaling analysis 
of seven properties; its value increases with the average 
valency. Scaling plots and data-collapse analyses show that the transition belongs 
to the standard percolation universality class even in disordered cases. 
\end{abstract}


\maketitle

\section{Introduction}\label{sec:intro}

Percolation~\cite{StaufferAharony94} is characterized by the formation of a 
spanning cluster in a system composed of elements~(sites and/or bonds) each 
present independently with probability $p$. The probability of a spanning or 
percolating cluster is only nonzero for $p > p_c$, the percolation threshold, 
marking a continuous phase transition with associated critical 
exponents~\cite{Fe80,Re58,Stauffer86}. Percolation has found a huge variety 
of applications, such as granular materials~\cite{Od98}, forest 
fires~\cite{He93}, polymers~\cite{Bu95,Fl41,Fl412,Fl413}, porous 
media~\cite{Ma91,Mo95}, and biological evolution~\cite{Ra94}.

Random sequential adsorption~(RSA)~\cite{Fe80,Ev92} is a stochastic process
consisting in irreversible deposition of immobile objects onto an initially
empty substrate, such that each object excludes a certain area from further
occupation. A realization of RSA stops when no further deposition events are
possible, at which point the system is said to be {\it jammed}.  Introducing a
deposition attempt rate~(per unit area, or per site, on a lattice) of unity,
a time can be associated with each deposition event in a given realization.
Letting the {\it coverage} $\rho$ denote the fraction of the substrate
occupied by deposited objects, we define $\rho\prt{t}$ as the mean coverage 
(over all realizations) at time $t$.  Of particular interest is the jamming 
or saturation coverage 
$\rho_\infty \equiv \lim_{t \to \infty} \rho\prt{t}$. RSA of objects of
diverse formats (discs~\cite{To10}, linear k-mers~\cite{Ko01,Bo94}, etc.)
has been used to model a wide range of physical  processes such as ion
implantation in semiconductors~\cite{Ro83}, protein adsorption~\cite{Fe80},
as well as the original car-parking problem~\cite{Ba93}. 

Consider RSA on a two-dimensional lattice, of objects occupying two or more
sites. As the coverage increases, it may be possible for the set of 
deposited objects to ``percolate," i.e., to form a spanning cluster, in 
which case the mean coverage at percolation is a quantity of interest.
(While RSA of monomers --- objects occupying a single site --- is trivial,
percolation of monomers is a classic problem and $p_c$ has been determined,
exactly or numerically, for a wide variety  of lattices~\cite{SykesEssam64,ScullardJacobsen2020,ZiffScullard06,RamirezCentresRamirezPastor19,SlutskiiBarashTarasevich18,MertensMoore2018}.)
RSA of extended objects, such as rods or linear lattice $k$-mers, is a
problem of current interest. Cherkasova~{\it et al.}~\cite{Ch10} showed that
the percolation threshold for dimers~($k$-mers with $k=2$) is smaller than
for monomers. As expected, the universality class does not depend on the length
$k$. In addition, these authors observed that when the dimers can only align
along one direction, the percolation threshold increases.

Cornette~{\it et al.}~\cite{Co03,Co032} studied the influence of $k$ in the
percolation threshold via Monte Carlo simulation~(MC) and Bethe lattice
analysis. They confirmed that $p_c$ decreases monotonically with $k$ for 
$1 \leq k \leq 16$, and that the universality class of the transition is standard
percolation, independent of $k$.  Leroyer and Pommiers~\cite{Le94}
demonstrated that the percolation threshold $p_c$ decreases with
increasing segment length, reaches a minimum value and then increases
for $k~\geq 15$.   
Tarasevich~{\it et al.}~\cite{Ta12,Ta15} conjectured, based on simulation,
that for $k$-mers of sufficient length~($k \gtrapprox 1.2\times 10^4$)  
percolation does not occur. More recently Kondrat~{\it et al.}~\cite{Ko17} 
developed a rigorous proof refuting this conjecture. 
They showed that for nonoverlaping $k$-mers, the jammed configuration includes a 
percolating cluster.

Although deposition of particles or $k$-mers on a regular
substrate has attracted much attention, this model needs to be extended to 
describe more realistic situations. In many
systems, the substrate includes impurities and/or defects that affect the
deposition process. A simple realization of substrate disorder involves
excluding randomly a fraction of sites from the deposition
process, resulting in a {\it diluted} system. For example,  
Cornette~{\it et al.}~\cite{co06} studied deposition of
polyatomic structures on diluted lattices, observing that
the percolation threshold increases with dilution, and that dilutions greater than 
a certain value, there is no percolation~\cite{co06,Ta15}. 

In standard dimer RSA, a given site can be
occupied by at most one monomer, and only one dimer can be incident
upon an occupied site.  A way of softening this restriction, while still
prohibiting arbitrarily high densities, is to allow up to $V_{max}$ dimers
to be incident on a given site.  In the resulting {\it restricted valence} 
RSA process, each lattice edge can be occupied by at most one dimer.
Here, $V_{max}$ ranges from unity (the usual dimer RSA problem) up to $z$,
the lattice coordination number.  (In the latter case, the RSA process is
trivial since all edges are eventually occupied.)  Restricted-valence RSA
is analogous to a self-avoiding random walk (SAW) in which the walk is 
allowed to visit the same site up 
to $K$ times. Krawczyk~\etal~\cite{Kr06} showed that existence and nature of
phase transitions in these models depends on the details and dimensionality.
For example, in case of forbidden reversal on the cubic lattice,
discontinuous and continuous transitions are observed, with the continuous
transition of the same type as in interacting self-avoiding walk
collapse~(ISAW)~\cite{Se90}. On the other hand, for allowed reversal on the
square lattice, these authors do not find any indication of a phase
transition.  Oliveira~\etal~\cite{Ol08} studied SAWs with $K=2$, using 
Husimi and Bethe lattice solutions. They found a rich phase diagram with
regular polymerized, nonpolymerized, and pair polymerized stable phases, a 
tricritical point and a critical endpoint. The transition between
polymerized and nonpolymerized phases can be continuous or discontinuous 
depending on the region of the phase diagram.

We consider restricted-valence RSA (and the associated percolation problem)
on the square and triangle lattices, in both its pure form (all sites having
the same maximum valence) and with disorder such that the the maximum
valences at each site are independent, identically distributed random variables. 
Our objective is to
understand how percolation thresholds and jamming limits depend on the 
valence restrictions. We verify that percolation transitions, when they
exist, belong to the standard percolation universality class. In addition,
we  develop a proof for the absence of percolation for maximum valence two.

The remainder of this paper is organized as follows. In Section
~\ref{sec:model} we detail the restricted-valence RSA model. Simulation
methods are presented in Sec.~\ref{sec:simdet} followed by results 
in Sec.~\ref{sec:resul}. Finally, in Sec.~\ref{sec:conclusions}, we present
our conclusions. The proof is given in the Appendix. 

\section{Model}\label{sec:model}

\begin{figure}[!htb]
    \centering
    \includegraphics[scale=.8]{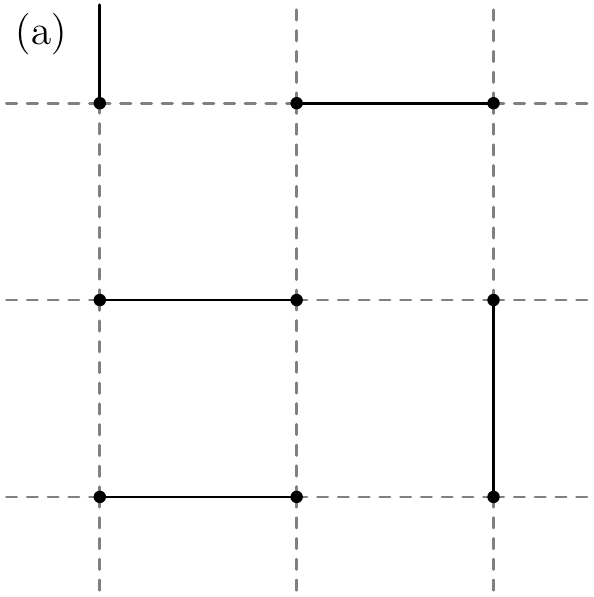}
    \includegraphics[scale=.8]{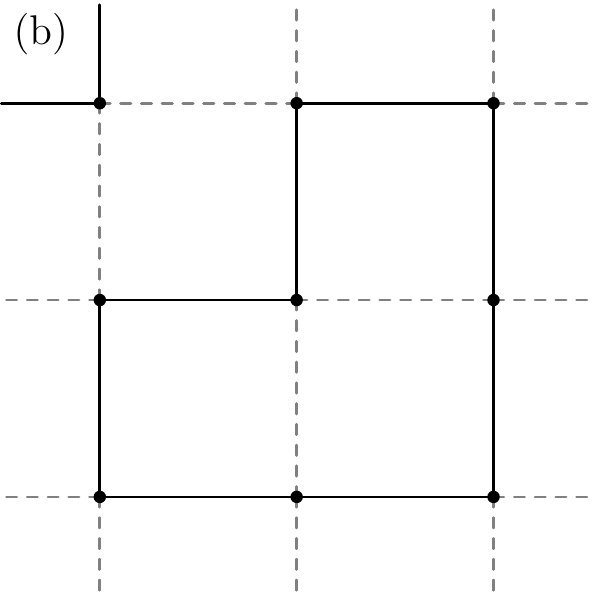}
    \includegraphics[scale=.8]{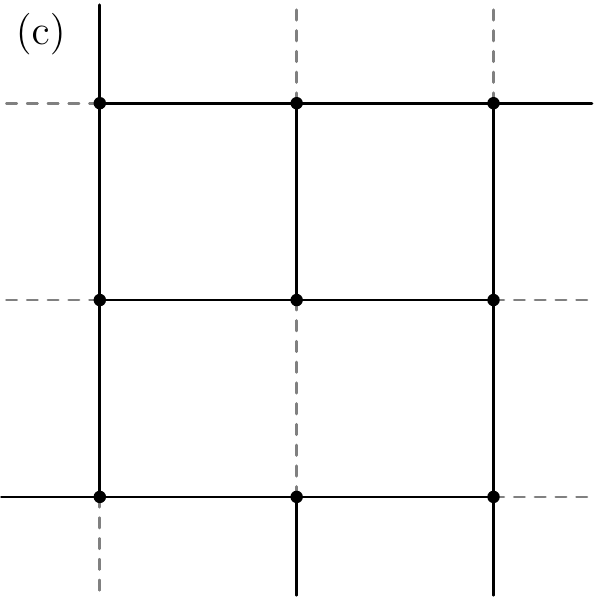}
    \caption{ Solid lines, dashed lines, and circles
     represent occupied edges, unoccupied edges and occupied sites 
    respectively. Panels (a), (b) and (c) 
    show possible configurations of systems with 
    $V_{\rm max}=1$, 2 and 3 respectively.}
    \label{fig:diags}
\end{figure}

We study RSA of dimers on a regular lattice under the restriction that the 
number of dimers that can attach to a vertex (its {\it valence}) cannot 
exceed $V_\mathrm{max}$. In case $V_{\rm max}=q$, with $q$ the coordination
number of the lattice, there is no restriction and all edges of the lattice
are eventually covered by a dimer. The cases $V_{\rm max}=1, 2$ and 3 
on the square lattice are shown in Figs.~\ref{fig:diags} (a), (b) and (c) 
respectively. For $V_\mathrm{max} = 1$, we have the usual irreversible 
dimer or domino tiling problem~\cite{Fl39}, in which only isolated edges 
may be occupied (see~~\fig~\ref{fig:diags}(a)). For $V_{\rm max}=2$, a 
vertex may have a maximum of two incident edges, giving rise to open or 
closed nonbranching paths, as shown in \fig~\ref{fig:diags}(b).
\fig~\ref{fig:diags}(c) shows a typical configuration for $V_{\rm max}=3$.
In~\fig~\ref{fig:diags} the bonds denoted by dashed lines cannot be
occupied, as this would violate the maximum-valence condition. Vertices 
with valence $V_{\rm max}$ are said to be {\it saturated}.

The RSA process is conveniently represented by associating times $t_i$ to
each edge $i$ of the lattice. The $t_i$, which are chosen anew at each realization 
of the process, are independent, identically distributed random variables, 
uniform on $(0,1]$.   At time zero, the lattice is empty (all edges unoccupied).  
At time $t_{min} \equiv \min_i \{t_i\}$, the edge corresponding to $t_{min}$ 
becomes occupied. Subsequently, edges are visited according to their associated 
times and occupied if this does not violate the maximum-valence conditions.  
Occupation is irreversible. Thus an unoccupied edge with one or more saturated 
vertices can never be occupied. Although percolating configurations exist for 
$V_{\rm max}=2$, the percolation probability is zero in this case, as shown in 
the Appendix.

In simulations on a periodic lattice of $N$ sites, there are $N_e = Nq/2$ edges.
The edges are ranked in order of increasing time and occupied (if permitted by the
maximum-valence restrictions) in that order.  We associate a discrete time 
$p_i = n_i/N_e$ with edge $i$, where $n_i \in \{1,...,N_e\}$ is the position
of edge $i$ on the ordered list.  Thus, for discrete time $p=0.5$, half the
edges have been visited in the RSA process. The process terminates when the
system is jammed, which happens for some (sample-dependent) $p_{\rm final} \leq 1$. 

\section{Simulation}\label{sec:simdet}

We use a variation of the union-find algorithm of Newman and 
Ziff~\cite{NewmanZiff00,NewmanZiff01} (NZ) to generate the RSA 
configurations efficiently and to estimate the percolation
point. In this algorithm, we first create a list of pairs of
neighboring vertices associated with each edge on the lattice.  For 
each realization of the RSA process, we generate a random ordering 
of this list.  In the usual NZ algorithm, we go down this list one 
pair at a time and add the bonds to the system, using the ``find" 
routine to find the roots of the two clusters at the ends of the 
new bond, and 
then the ``union" step to join two clusters if their roots are 
currently marked as distinct. Ordering the bonds beforehand is very
useful for the RSA problem since we can just go down this list, thus
considering each bond just once.  For the restricted-valence
models, we modify this program to keep track of the valence of each 
site, and only occupy a bond if the valences of its two sites are 
$< V_\mathrm{max}$. 

We study lattices of linear size $L=$32, 48, 64, 96, 128, 192, 256, 384,
512 and 768 corresponding to $N =(q/2)L^2$ bonds, where $q=4$ (6) for 
square (triangular) lattices. To estimate the properties of interest 
we average over $10^6$ independent realizations 
starting from an empty lattice.

\section{Results}\label{sec:resul}

In this section we report results the for the percolation threshold, 
critical exponents and jamming coverage for restricted-valence RSA 
on the square and triangle lattices. The properties of interest are 
exhibited as functions of the control parameter $p$, i.e., the fraction of bonds 
visited. In percolation, the order parameter is usually defined 
as~\cite{Ja19} 
\begin{equation}\label{eq:ordpar}
\Omega\prt{p,L} \equiv \aver{s_{\rm max}}/L^d,
\end{equation}
where $s_{\rm max}$ is the mean fraction of sites in the largest cluster, 
$d$ is the dimensionality, and
the angular brackets denote an average over realizations.
The percolation threshold of the infinite lattice
$p^\infty_{c}$ is estimated via finite-size scaling~(FSS)
analysis~\cite{StaufferAharony94} and is expected to follow,
 \begin{equation}\label{eq:scalpcgen}
    p^\infty_c - p_c\prt{L} \sim L^{-1/\nu},
\end{equation}
where $p_c\prt{L}$ is the {\it pseudocritical} value for lattice size $L$ 
and $\nu$ is the critical exponent governing the correlation length. The 
pseudocritical value is commonly determined through the position of the 
maximum of some ``diverging" quantity, or of a crossing point for different 
system sizes. In this work we analyze a set of seven quantities; five are 
moment ratios of the form,

\begin{equation}\label{eq:momrat}
    Q^{js}_{ir}\prt{p,L} \equiv 
    \frac{M_n\prt{p,L}}{M^i_r\prt{p,L} M^j_s\prt{p,L}} \quad 
    {\rm with} \quad ir+js = n,
\end{equation}

where the moments $M_k$ are defined as 
$M_k\prt{p,L} \equiv \aver{s^k_{\rm max}}$ and $k=1,\ldots,4$,
\begin{equation*}
    Q^{11}_{11} = \frac{M_2}{M^2_1}, \quad 
    Q^{\frac{1}{2}2}_{12} = \frac{M_3}{M^{3/2}_2}, \quad     
    Q^{12}_{12} = \frac{M_4}{M^2_2}, \quad 
    Q^{11}_{21} = \frac{M_3}{M^3_1}, \quad 
    Q^{11}_{12} = \frac{M_3}{M_1M_2}.
\end{equation*}
We also study the second cumulant $K_2(p,L)$,
\begin{equation}
    K_2\prt{p,L} = L^{-2}\prt{M_2 - M^2_1} 
\end{equation}
This property, also called the susceptibility $\chi\prt{p,L}$,
provides information about the fluctuations in the size of the largest 
cluster. Finally, we analyze $M^\prime_2\prt{p,L}$, defined as the 
difference between the second moment of s $M_2\prt{p,L}$ and the mean 
fraction of sites in the largest cluster:
\begin{equation}\label{eq:m2p}
    M^\prime_2\prt{p,L} = \frac{1}{L^d}
    \sum_{i\neq {\rm max}}s^2_i = 
     M_2\prt{p,L} - \frac{\aver{s^2_{\rm max}}}{L^2},
\end{equation}
where 
\begin{equation}\label{eq:m2}
    M_2\prt{p,L} =\sum_{s}s^2n_s = \frac{1}{L^2}\sum_i s_i^2.
\end{equation}
In Eqs.~(\ref{eq:m2p}) and~(\ref{eq:m2}), $s$ denotes cluster size and $n_s$ the average number of clusters of size $s$. 

The FSS theory of percolating systems~\cite{Ta87} states that at the
critical point $p=p^\infty_c$, $\Omega\prt{p,L}$, $M^\prime_2\prt{p,L}$ 
and $K_2\prt{p,L}$ obey the relations, 
\begin{equation}\label{eq:beta}
    \Omega\prt{p=p^\infty_c,L} \sim L^{-\beta/\nu}
    {\cal U}\prt{\varepsilon L^{1/\nu}},
\end{equation}
\begin{equation}\label{eq:gamma1}
    M^\prime_2\prt{p=p^\infty_c,L} \sim L^{-\gamma/\nu}
    {\cal M}\prt{\varepsilon L^{1/\nu}},
\end{equation}
and
\begin{equation}\label{eq:gamma2}
    K_2\prt{p=p^\infty_c,L} \sim L^{-\gamma/\nu}
    {\cal K}\prt{\varepsilon L^{1/\nu}}
\end{equation}
where $\beta,\gamma$ and $\nu$ are critical exponents
and $\varepsilon \equiv {p-p_c}$ represents the distance to
the critical point. The scaling functions ${\cal U}(x)$, ${\cal M}(x)$ and 
${\cal K}(x)$ exhibit universal behavior.

\subsection{Square lattice}\label{sec:sqlatresul}
\subsubsection{Determination of \texorpdfstring{$p^\infty_c$}{Lg}}
As mentioned in Sec.~\ref{sec:model}, for maximum valence $V_{\rm max}=1$ 
on the square lattice, there is no percolation. For $V_{\rm max}=2$ a
typical jammed configuration is shown in \fig~\ref{fig:snapshot}(a).
In this case there is no percolation, as shown in the Appendix. 
On the other hand, for $V_{\rm max}=3$, see \fig~\ref{fig:snapshot}(b),
there are large connected regions.

\begin{figure}[!htb]
    \includegraphics[width=6cm, height=6cm]{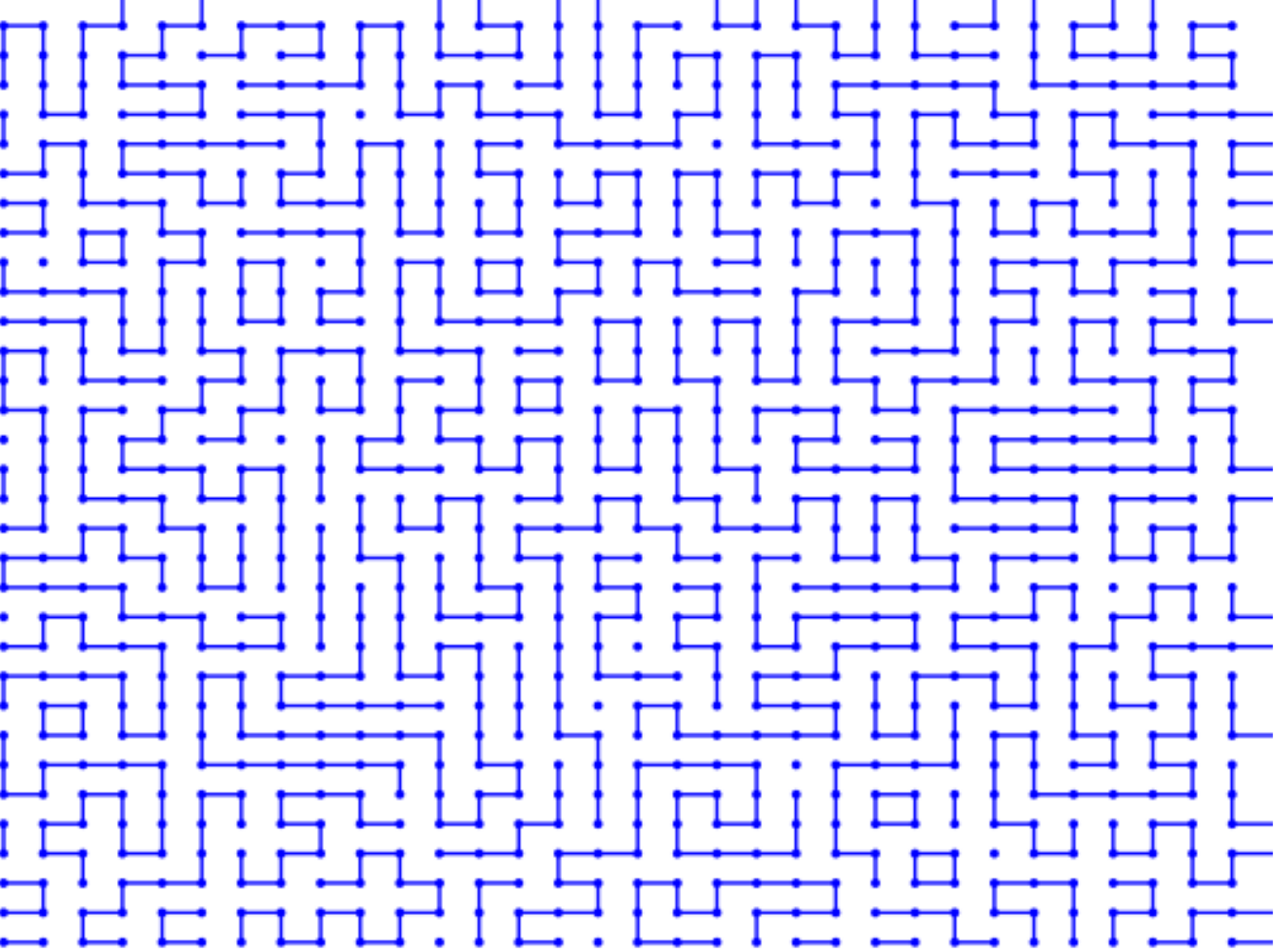}
    \includegraphics[width=6cm, height=6cm]{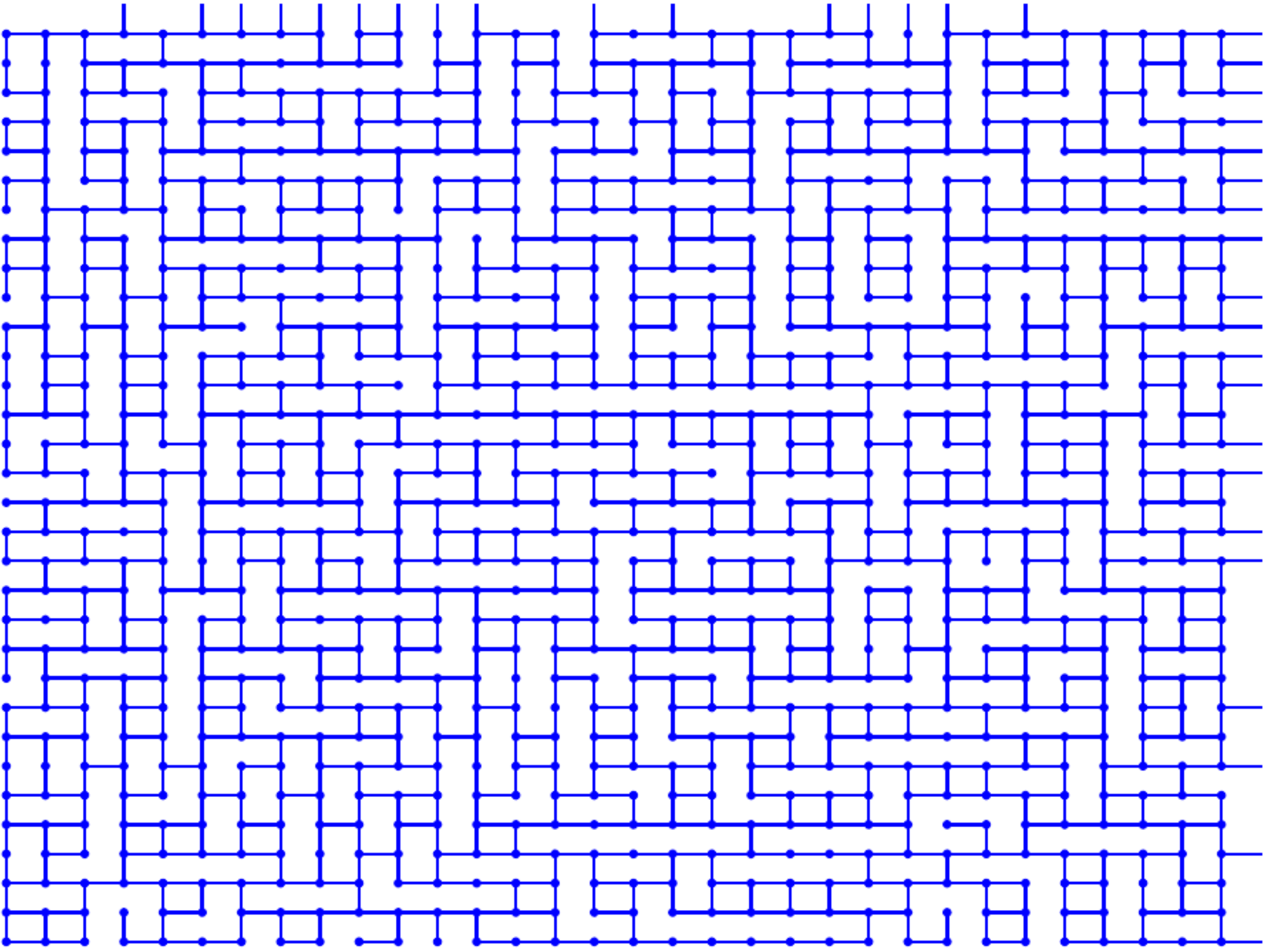}
  
   (a) \hspace{5cm} (b) \\ 
    \caption{Typical jammed configurations on a $32 \times 32$ lattice for the  
    $V_{\rm max}=2$(a) and $V_{\rm max}=3$(b).}    
    \label{fig:snapshot}
\end{figure}

\begin{figure}[!htb] 
   \centering
    \includegraphics[scale=2.]{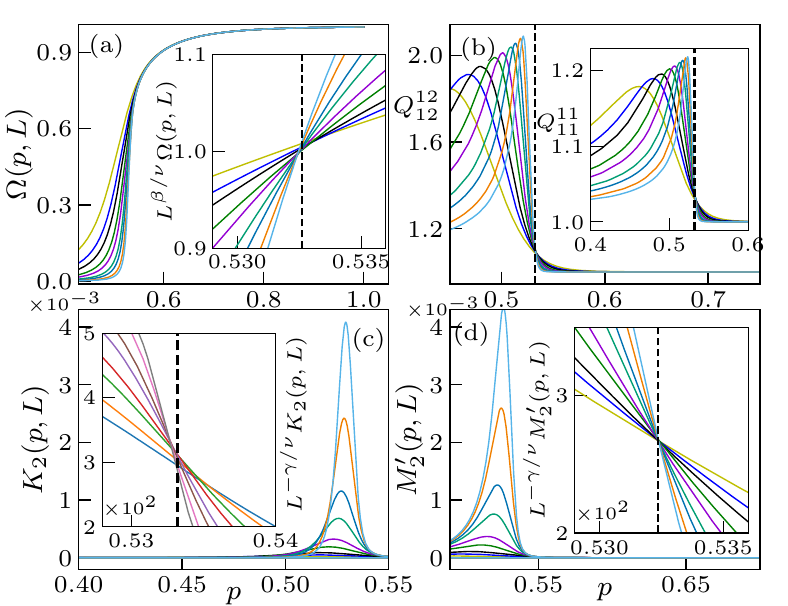}
   \caption{Results for $V_{\rm max}=3$. (a) Main graph:
   order parameter~$\Omega(p,L)$ versus $p$, with system 
   size increasing from left to right. Inset: 
   $L^{\beta/\nu}\Omega$ versus $p$ near the crossing point (vertical line). 
   (b) Main~(inset) 
   moment ratio $Q^{12}_{12}$~($Q^{11}_{11}$) 
   versus $p$ with system size increasing from left to 
   right. (c) Main graph: $K^\prime_2$; inset: detail of the crossing 
   region of the scaled $K^\prime_2$ with system size increasing 
   from bottom to top. The dashed line marks the crossing point. 
   (d) Similar to (c), but for $M^\prime_2$.}
   \label{fig:allsq}
\end{figure}

Results for the properties mentioned
above~(for $V_{\rm max}=3$) are shown in~\fig~\ref{fig:allsq}. The abrupt increase of
$\Omega\prt{p,L}$~(\fig~\ref{fig:allsq}(a)) suggests a phase transition 
at some value of $p$ between 0.40 and 0.60. The inset of this figure
shows an expanded plot of the order parameter multiplied by the factor
$L^{\beta/\nu}$, where $\beta=5/36$ and $\nu = 4/3$ correspond to the 
standard two-dimensional percolation critical exponents. A crossing point is evident at
$p\approx 0.532$. Figure~\ref{fig:allsq}(b), which shows the moment 
ratio $Q^{12}_{12}$, again suggests a continuous transition, 
with a crossing point for $p \approx 0.53$. The 
inset of panel~(b) shows that $Q^{11}_{11}$ exhibits similar behavior. The 
moment ratios exhibit strong finite-size effects, with the pseudo-critical 
values of $Q$ increasing substantially with system size.
Figure~\ref{fig:allsq}(c) shows the second cumulant $K_2$, which also 
exhibits signatures of a continuous transition. The peaks occur in 
the range $0.51<p<0.54$ in agreement with the other properties. The 
inset of this figure shows the second cumulant scaled by 
$L^{\gamma/\nu}$, with $\gamma=43/18$ the critical exponent that 
governs the fluctuations in the largest cluster size. Although we 
observe some discrepancies for $L=32$ and $48$, the curves for larger 
systems intersect at $p\approx 0.532$. $M^\prime_2$, plotted in
\fig~\ref{fig:allsq}(d) follows the same tendencies as the other 
properties, exhibiting maxima that increase systematically with 
system size. The peaks appear in the range $0.50 < p < 0.54$. The 
scaling plot~(inset) shows a crossing at $p=0.532$

\begin{figure}[!htb]
\includegraphics[scale=1.5]{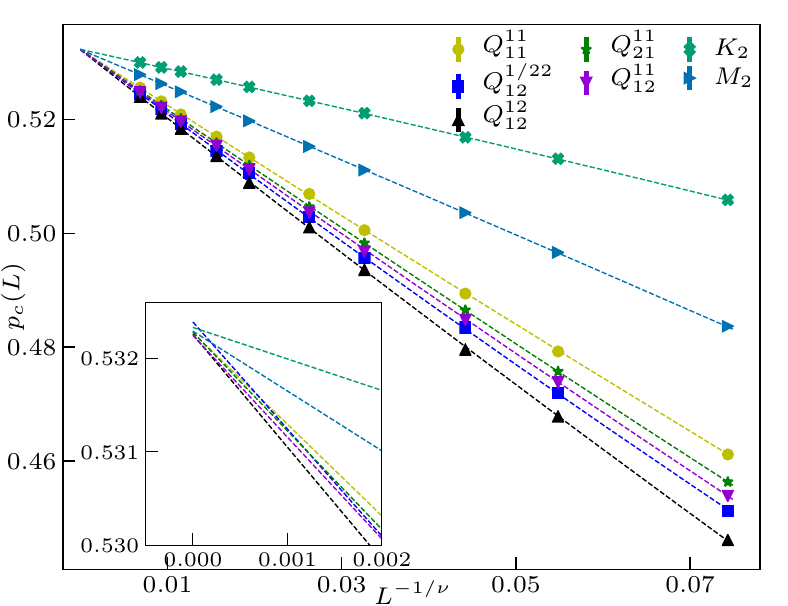}
\caption{FSS of the pseudocritical points. Symbols represent
simulation values and lines linear fits.  Error bars are smaller than 
symbols. Inset: detail of the limiting ($L \to \infty$) region.}
\label{fig:fsspc}
\end{figure}

FSS analysis of the pseudocritical points, shown
in~\fig~\ref{fig:fsspc}, reveals that for each property, $p_c(L)$ 
is well fit by a straight line when plotted versus $L^{-1/\nu}$. Here the 
pseudocritical points $p_c$ for each quantity are estimated using a 
polynomial fit to approximately 10 points around the global maximum. (The
order of the polynomial is chosen as the lowest order that yields residuals
without systematic behavior. In most cases fourth-order polynomials are
used). The uncertainty in the position of the maximum was estimated through
the root mean square deviation~(RMSD); in the worst case, the RMSD 
$\sim 10^{-4}$~({see~\tab~\ref{tab:pseudovals}}).

\begin{table*}[!htb]
    \centering
    \begin{tabular}{r|c|c|c|c|c|c|c}
    \hline \hline
    $L$ & $Q^{11}_{11}$ & $Q^{12}_{12}$ & 
    $Q^{\frac{1}{2}2}_{12}$ &  $Q^{11}_{21}$ & $Q^{11}_{12}$ &
    $K_2$ & $M^\prime_2$ \\ \hline 
    32 & 0.46112(1) & 0.4460(1)  & 0.45122(2) &  0.45629(5) &
    0.45385(3) & 0.5058(1) & 0.48366(4) \\ 
    48 & 0.47923(1) & 0.4677(1)  & 0.47188(4) & 0.4757(1) & 
    0.47381(6) & 0.51310(4) & 0.49663(4)  \\
    64 & 0.489401(9) &  0.4795(1) & 0.48343(2) & 0.48650(6) &
    0.48477(4) & 0.51687(4) & 0.50362(2) \\
    96 & 0.500547(6) & 0.4935(1) & 0.49565(1) & 0.49828(4) &
    0.49678(4) & 0.52112(4) &  0.51112(1) \\
    128 & 0.506917(6)  & 0.50105(8) & 0.50287(1) & 0.50466(4) &
     0.50366(3) & 0.52327(2) & 0.51526(1)  \\
    192 & 0.513341(6) & 0.50887(8) & 0.51066(1) & 0.51192(3) & 
     0.51126(2) & 0.52572(2) & 0.51976(1)  \\
    256 & 0.516989(8) & 0.51359(5) & 0.51446(1) & 0.51588(3) &
    0.51541(1) & 0.52700(3)  & 0.522269(9) \\
    384 & 0.520888(3) & 0.51837(4) & 0.519221(5) & 0.52015(2) &
     0.51964(1) & 0.52842(1) & 0.524864(8) \\
    512 & 0.523199(3)  & 0.52104(5) & 0.52177(1) & 0.52248(2) & 
    0.52207(1) & 0.52916(1) & 0.526303(8) \\
    768 & 0.525601(4)  & 0.52400(3) & 0.524578(7) & 0.525048(1) &  
    0.52478(1) & 0.53003(1) & 0.527896(6) \\ \hline
    $\infty$ & 0.5322(1) & 0.5323(1) & 0.5322(1) & 0.53228(4) & 
    0.5322(1) & 0.53233(3) & 0.53222(5)  \\  \hline \hline 
    \end{tabular}
    \caption{Estimates for the pseudocritical points for each
    system size $L$ and quantity analyzed. The final line reports
    extrapolated ($L \to \infty$) estimates for the critical point. 
    Numbers in parentheses denote uncertainties.}
   \label{tab:pseudovals}
\end{table*}

In the limit $L \to \infty$ all the estimates for the percolation
threshold converge to very similar values~(see
Fig.~\ref{fig:fsspc} and~\tab~\ref{tab:pseudovals}). The final 
estimate for the percolation threshold, $p^\infty_c$ is
obtained through a weighted average of the estimates associated with 
each property, with weights $\propto 1/\sigma^2$, where $\sigma$ 
represents the uncertainty of each estimate, yielding a final estimate
of $p^\infty_c=0.5323(1)$ for $V_{\rm max}=3$ on the square lattice.

\begin{figure}[!htb]
    \centering
    \includegraphics[scale=1.5]{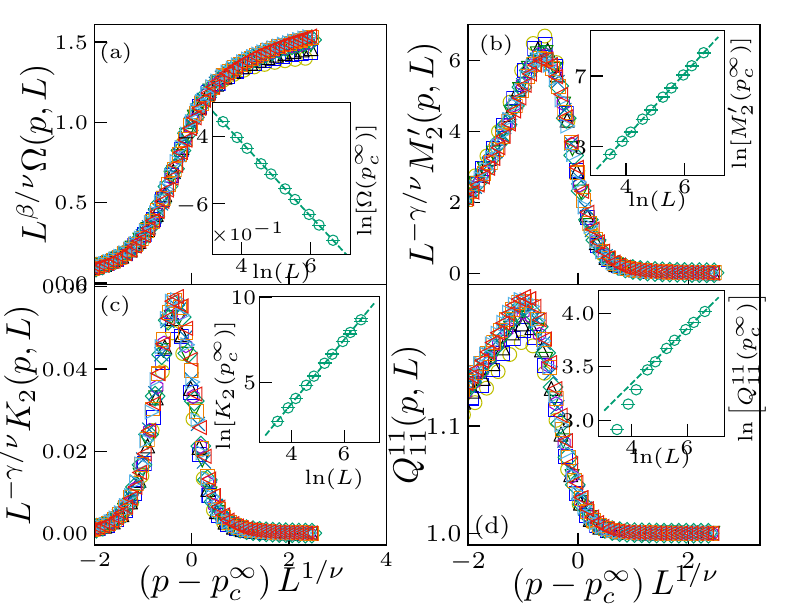}
    \caption{Panels (a), (b), (c) and (d) exhibit data 
    collapses for $\Omega$, $M^\prime_2$, $K_2$ and $Q^{11}_{11}$, 
    respectively. Insets are log-log plots of the aforementioned 
    properties at the critical point. Exact critical exponent
    values for standard percolation in two dimensions are 
    used in the data collapse analysis. Symbols 
    $\circ$, $\square$, $\triangle$, $\triangledown$, $\hexagon$,
    $\lozenge$, $\times$, $\square$, $\triangleright$,
    $\triangleleft$ correspond to system sizes 
    in ascending order.}
    \label{fig:collapse}
\end{figure}

The results for the scaled values of $\Omega$, $K_2$ and $M^\prime_2$ 
suggest that the transition belongs to the standard percolation universality class (SPUC).
We verify this
conclusion via data-collapse analyses, as shown in~\fig~\ref{fig:collapse}. 
Panel (a) shows that the SPUC critical exponents yield a good collapse of 
the simulation data for different system sizes. The log-log plot of the 
order parameter (inset (a)) provides $\beta/\nu=0.109(3)$ quite close 
to the exact value, $\beta/\nu=0.104166\ldots$~\cite{StaufferAharony94}, 
for SPUC in two dimensions. The data collapse of $M^\prime_2$~(see panel 
(b)) also exhibits a good overlap, although slight deviations 
are observed for $L=32$ and $L=48$ in the vicinity of the critical
point. The log-log plots provide a slope of 1.800(1), in 
agreement with the theoretical prediction of the ratio $\gamma/\nu$ 
for SPUC. For $K_2$~(see \fig~\ref{fig:collapse}(c)), the data 
points and the log-log plots~(see inset) provide a slope of 
1.782(8), again in agreement with the exact result 
$\gamma/\nu=1.7916\ldots$~\cite{StaufferAharony94}. The ratios $\beta/\nu$ and $\gamma/\nu$ satisfy the 
hyperscaling relation $d\nu = \gamma + 2\beta$ within uncertainty. 

\subsubsection{Valence disorder}

We now analyze restricted-valence RSA with valence disorder, in which
the valences associated with each site are taken as independent, 
identically distributed random variables. We consider valence 
distributions uniform on the set 
$\left\{V_{\rm i},\ldots,V_{\rm j}\right\}$.  At 
each realization, a new set of valences is generated. On the square 
lattice, we considered $\prt{V_{\rm i},V_{\rm j}}=$ $\prt{1,4},$ 
$\prt{1,3},$ $\prt{1,2},$ $\prt{2,4},$ $\prt{2,3}$ and $\prt{3,4}$, with each valency $V_{\rm i}$ occurring with 50\% probability.

\begin{figure}[!htb]
    \centering
    \includegraphics[scale=1.8]{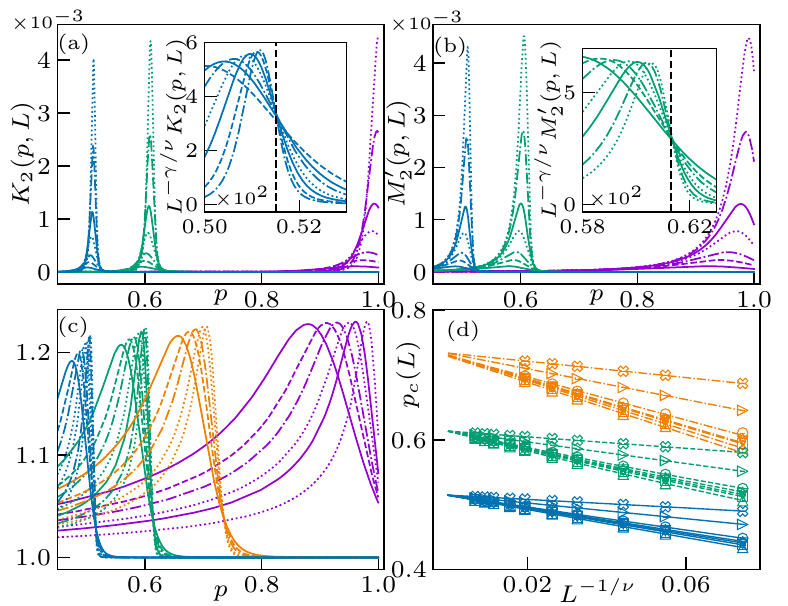}
    \caption{(a) $K_2(p,L)$ versus concentration $p$. Different 
    line patterns denote different system sizes while the colors
    blue, green and magenta correspond to 
    $\prt{V_{\rm i},V_{\rm j}}=(3,4), (2,4)$ and (1,4)
    respectively. The case (2,3) (not shown) exhibits similar
    behavior. The inset shows the crossing points of $K_2$ scaled 
    using standard percolation exponents. (b) is a analogous to (a), but
    for $M_2\prt{p,L}$. (c) $Q^{11}_{11}$ versus $p$ for cases
    (2,3), (3,4), (2,4) and (1,4). (d) Pseudocritical points versus 
    $L^{1/\nu}$.} 
    \label{fig:distm2xi}
\end{figure}

There is no percolation for cases (1,3) and (1,4).  
While disorder alters the percolation thresholds and jamming coverages, it does 
not appear to affect the critical exponents. According to the Harris 
criterion~\cite{Ha74,Be93}, disorder is relevant if $d\nu \leq 2$. Since 
the transition in the pure system belongs to the SPUC with 
$d\nu = 2(4/3) > 2$, disorder is indeed expected to be irrelevant. The 
quantities $K_2\prt{p,L}$, $M^\prime_2\prt{p,L}$ and $Q^{11}_{11}$ in 
the presence of disorder are shown in Figs.~\ref{fig:distm2xi}(a), (b) 
and (c) respectively. 

Figure \ref{fig:distm2xi} shows clear evidence of a continuous 
percolation transition, as in the pure case. The insets
of Figs.~\ref{fig:distm2xi}(a) and (b) exhibit crossings using the 
standard percolation critical exponents, suggesting that disorder 
does not affect the universality class of the transition. 
Note as well that the crossing value of $Q^{11}_{11}$ is the
same as for the pure case.
As expected, the percolation threshold moves to lower values of $p$ 
as the mean valence increases. For example, for $(3,4)$ (blue 
lines in~\fig~\ref{fig:distm2xi}(a), (b) and (c)), half the sites have 
maximum valence three and half maximum valence four. Both 
valences percolate and the percolation threshold occurs at a value 
between those for the pure cases. 

Our estimates for the percolation threshold are obtained via FSS as 
shown in~\fig~\ref{fig:distm2xi}(d), providing
$p^\infty_c=0.7283(5)$, $0.6099(1)$, $0.5120(2)$ for the cases 
$(2,3)$, $(2,4)$ and $(3,4)$ respectively. The analyses follow the 
same lines as the pure cases $\prt{3,3}$ and $\prt{4,4}$.
It is worth noting that although (1,4) and (2,3) possess the same mean valence,
the latter percolates while the former does not.

\subsubsection{Coverage density and jamming state}

Without disorder, case $(1,1)$ is the most restrictive, hence 
the lowest jamming coverage is obtained. In this case each 
absorbed bond prevents all its neighbor bonds from being occupied, 
so that in a perfectly regular pattern, corresponding to full 
coverage of sites, only 1/4 of the bonds are occupied. Analysis of this 
case goes back to the work of Nord and Evans~\cite{No85},
who obtained a saturation coverage $0.9068$. Subsequently, 
de Oliveira {\it et al.}~\cite{Ol92} obtained $0.90677(6)$ via 
series expansions and simulation. Our approach provides 
0.906814(5)~(a deviation of less than $0.0045\%$ compared 
with~\cite{Ol92}). For~$(3,3)$ the insertion of a bond blocks 
two of its six neighbors and thus, in a regular 
pattern, $3/4$ of bonds can be occupied. In this case we find a 
coverage of $0.999391(1)$. Case $(4,4)$ is trivial since all
edges are occupied by bonds. 

\begin{figure}[!htb]
    \centering
    \includegraphics[scale=.9]{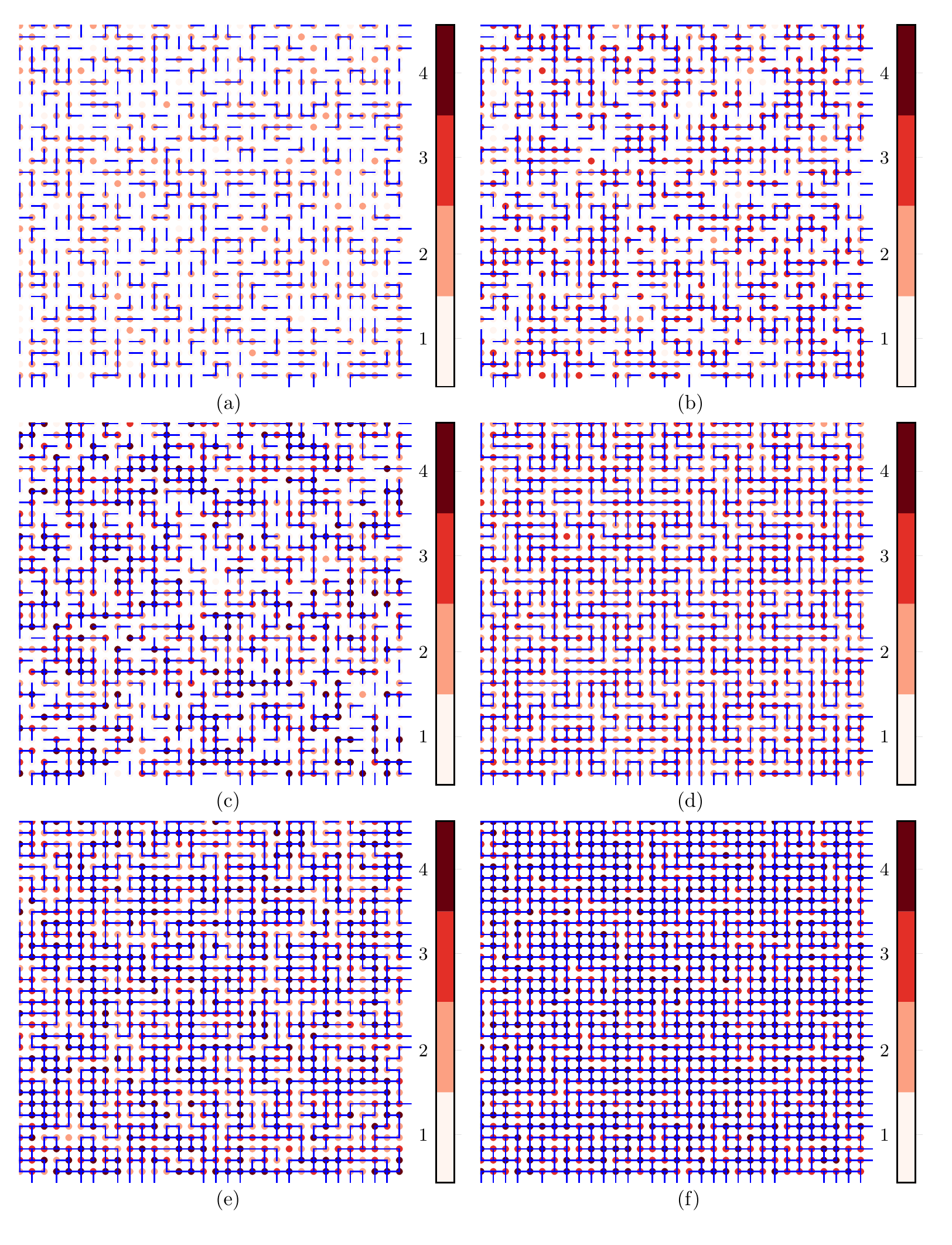}
    \caption{Typical jammed configurations for disordered 
    cases $(1,2)$, $(1,3)$, $(1,4)$, $(2,3)$, $(2,4)$ and 
    $(3,4)$~(panels a-f, respectively) for $L=32$. Color scales 
    denote valences. Blue lines denote occupied bonds.}
    \label{fig:snapsdis}
\end{figure}

Typical jammed configurations for disordered cases are shown  
in~\fig~\ref{fig:snapsdis}. For~$(1,2)$ 
(\fig~\ref{fig:snapsdis}(a)), only sites with valence 1 and 2 
are permitted. Since half the sites have valence one, there are 
isolated bonds and terminal points. 
Figure 7(c) shows (1,4), which is the most heterogeneous.  All possibilities are observed in the snapshot, from isolated bonds to sites with the maximum number of bonds.  Case (3,4) (Fig. 7(f)) has the highest density of bonds; nevertheless there are sites with 
valence 4 that have only two incident bonds.

\begin{figure}[!htb]
    \centering
    \includegraphics[scale=1.8]{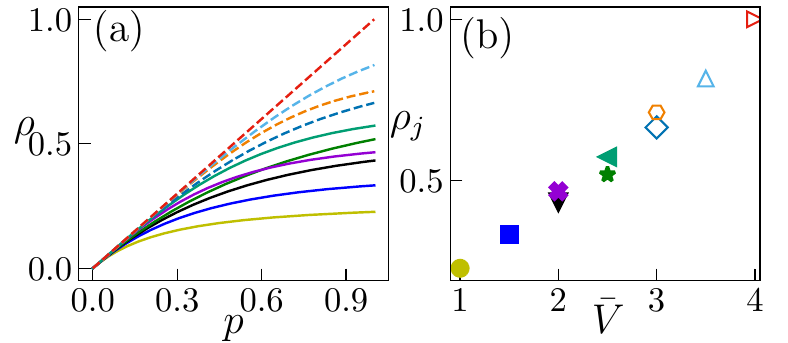}
    \caption{(a)~Coverage density $\rho$ $versus$ $p$ 
    for $L=128$ and distinct combinations of valencies. From 
    bottom to top, (1,1), (1,2), (1,3), (1,4), (2,2), (2,3)
    (solid lines) and (2,4), (3,3), (3,4) and (4,4) (dashed lines). 
    (d)~The filled~(opened) symbols represent the jamming fraction 
    as a function of the average valence. Filled~(opened) symbols 
    correspond to the solid~(dashed) lines in the plot~(a).}
    \label{fig:covdenssq}
\end{figure}

Results for the coverage density $\rho$ as a function of $p$ and 
the jamming coverage are shown in 
\fig~\ref{fig:covdenssq}. Figure~\ref{fig:covdenssq}(a) exhibits 
the density of occupied bonds as a function of $p$. The red dashed 
line represents the limiting case $(4,4)$ in which all bonds are 
occupied. While the curves are generally similar, those for $(1,4)$ 
and $(2,2)$ cross near $p=0.60$, suggesting that the rate at 
which bonds are occupied changes in a nontrivial manner as $p$ varies. 
In \fig~\ref{fig:covdenssq}(d) we plot the jamming density $versus$ 
$\bar{V}$, the arithmetic average of the valencies in the range 
$\{V_{i},\ldots,V_{j}\}$; $p_j$ increases approximately 
linearly with $\bar{V}$. 

\subsubsection{Percolation thresholds}
We determine the bond occupation fractions at the pseudocritical percolation 
points, $\rho_c(L)$, which, when extrapolated to infinite system size, furnish 
estimates for the bond percolation densities, $\rho_c$, for the different cases 
of pure and mixed valences (see Table~\ref{tab:sq}).  Pure valence-4 is simply 
bond percolation on the square lattice and our result is consistent with the exact 
value, $\rho_c =1/2$~\cite{Stauffer86}. At the other extreme, (2,3), $\rho_c$ is 
clearly larger than one-half. Recalling that the pure valence-2 system does not 
percolate, $\rho_c > 1/2$ can be understood qualitatively by noting that many 
occupied bonds falling in regions rich in valence-2 sites cannot contribute to 
percolation.  A similar observation applies to the (2,4) mixture.  For (3,3) the 
deviation from $\rho_c = 1/2$ is barely significant, while for (3,4) our result is 
consistent with a percolation density of $1/2$. It is nonetheless surprising that 
the percolation density for (3,3) is {\it smaller} than 1/2.  While we 
defer a detailed study of this case to future work, we note that in all cases except 
(4,4), the occupation of bonds incident upon a given site are not independent events, 
so that deviations from standard independent percolation are possible in principle.

We summarize our results for the percolation threshold, jamming coverage and the 
density of occupied sites at jamming for the square lattice in~\tab~\ref{tab:sq}.
\begin{table}[!htb]
\begin{tabular}{c|c|c|c|c}
\hline\hline
    $(V_{i},V_{j})$ & $p^\infty_c$ & $\rho_j$ & 
    $\theta_{\rm sites}$ & $\rho^\infty_c$ \\ \hline
    $(1,1)$ & -----     & 0.226705(3) & 0.906814(5)  & ----- \\
    $(1,2)$ & -----     & 0.332928(2)  & 0.958746(1) & ----- \\
    $(1,3)$ & -----     & 0.432512(6) & 0.982236(1)  & -----\\
    $(1,4)$ & -----     & 0.517817(6) & 0.992235(2)  & -----\\
    $(2,2)$ & -----     & 0.465780(5) & 0.987851(9)  & -----\\
    $(2,3)$ & 0.7283(5) & 0.572630(2) & 0.995678(9) & 0.512(2) \\
    $(2,4)$ & 0.6099(1) & 0.664237(1) & 0.998557(9) & 0.5070(6)\\
    $(3,3)$ & 0.5323(1) & 0.710935(1) & 0.999391(1) & 0.496(2) \\
    $(3,4)$ & 0.5120(2) & 0.816074(5) & 0.999903(3) & 0.498(3) \\
    $(4,4)$ & 0.5000(1) & 1.00000000 & 1.00000000   & 0.5000(1)\\
    \hline\hline
\end{tabular}
\caption{Simulation results for $p^\infty_c,p_j$,
$\theta_{\rm sites}$ and $\rho^\infty_c$ for the square lattice. 
The horizontal lines in the first column correspond to cases 
without percolation.} 
\label{tab:sq}
\end{table}

\newpage

\subsection{Triangle lattice}

We adopt the same approach as employed for the square lattice; the 
scaling behaviors are quite similar, hence we only report numerical values. 
For $(1,1)$, $(2,2)$, $(1,2)$ and $(1,3)$, valence restrictions prohibit
percolation. All other cases exhibit a percolation transition 
characterized by size-dependent peaks in $M^\prime_2$, $K_2$, and the
$Q^{js}_{ir}$. Table~\ref{tab:pstri} exhibits
simulation values for case $(4,4)$.

\begin{table}[htb]
\begin{tabular}{c|c|c|c|c|c|c|c}
\hline\hline
   $L$ & $Q^{11}_{11}$ & $Q^{12}_{12}$ & 
    $Q^{\frac{1}{2}2}_{12}$ &  $Q^{11}_{21}$ & $Q^{11}_{12}$ &
    $K_2$ & $M^\prime_2$ \\ \hline  
  32  & 0.30009(3) & 0.2887(1) & 0.29211(4) & 0.297392(4) & 0.29445(6) &
  0.33240(2) & 0.319581(2) \\
48  & 0.31363(2) & 0.3050(2) & 0.30789(4) & 0.307626(2) & 0.30875(7) & 0.33753(9) & 0.328151(1) \\
64  & 0.32089(2) & 0.3148(1) & 0.31601(2) & 0.318307(1) & 0.31754(7) & 0.34056(3) & 0.3311319(5) \\
96  & 0.32892(2) & 0.3234(2) & 0.32586(3) & 0.327910(1) & 0.32617(5) & 0.34345(6) & 0.3367670(4) \\
128 &  0.33325(1) & 0.3292(1) & 0.3301(4) & 0.332074(1) & 0.33126(4) & 0.34520(1) & 0.3393849(7) \\
192 &  0.33817(1) & 0.3350(1) & 0.3361(4) & 0.336908(1) & 0.33636(3) & 0.34706(1) & 0.3425698(7) \\
256 & 0.34082(1) & 0.33839(9) & 0.3393(3) & 0.340100(1) & 0.33937(4) & 0.34795(1) & 0.3445013(4) \\
384 & 0.34364(1) & 0.34190(8) & 0.3424(1) & 0.3429443(9) & 0.34273(2) & 0.34903(9) & 0.3464356(3) \\
512 &0.34528(1) & 0.34375(7) & 0.34433(1) & 0.3447238(7) & 0.34447(1) & 0.34964(2) & 0.3475288(7) \\
768 & 0.34705(1) & 0.34584(4) & 0.3462(1) & 0.346731(3) & 0.346494(1) & 0.35030(7) & 0.348957(9) \\
\hline 
$\infty$ &  0.3518(5) & 0.3516(3) & 0.3518(5) & 0.35172(1) & 0.35177(7) & 0.3519(7) & 0.35175(1)\\
\hline \hline
\end{tabular}
\caption{Estimates for the pseudocritical points and other quantities for each system size $L$ for (4,4). The final line reports extrapolated ($L \to \infty$) estimates for critical values. Numbers in parentheses denote uncertainties.}
\label{tab:pstri}
\end{table}

In the limiting case $(6,6)$, the percolation threshold 
is known exactly: $p_c = 2\sin(\pi/18)=0.347296355...$~\cite{SykesEssam64}. 
Our estimate of $p^\infty_c(6,6)=0.3472(5)$ is in agreement with this result.  
Estimates for the other cases are shown in \tab~\ref{tab:pcpj}. Analysis of 
critical exponents again yields values consistent with SPUC. For example, 
scaling plots of $M^\prime_2$ and $\Omega$ using SPUC exponents exhibit 
crossings of curves at the percolation threshold.

The jamming probability and coverage density are determined via 
the same techniques used for the square lattice. The curves for the density 
of occupied bonds for $(2,5)$ and $(2,6)$ cross the curve for $(3,3)$, similar
to the cases $(1,4)$ and $(2,2)$ on the square lattice. A compendium of
results for the triangle lattice is reported in~\tab~\ref{tab:pcpj}.  Different from
the square lattice, here all percolation thresholds are greater than or equal to the
exact value for the unrestricted triangle lattice.

\begin{table}[!htb]
   \centering
    \begin{tabular}{c|c|c|c|c}
     \hline\hline
      $(V_{i},V_{j})$ & $p^\infty_c$ & $\rho_j$ & $\theta_\mathrm{sites}$ & 
      $\rho^\infty_c$ \\
          \hline
      (1,1)  & -----     & 0.152338(1) & 0.914028(2) &\\
      (1,2)  & -----     & 0.229197(3) & 0.966224(2) & \\
      (1,3)  & -----     & 0.302104(5) & 0.987364(7) & \\
      (1,4)  & 0.8582(5) & 0.371539(2) & 0.995508(5) & 0.35(3) \\
      (1,5)  & 0.6374(5) & 0.434392(1) & 0.998354(5) & 0.35(1)\\
      (1,6)  & 0.5489(4) & 0.488033(5) & 0.999309(9) & 0.361(3)\\
      (2,2)  & -----     & 0.312030(7) & 0.983257(6) & \\
      (2,3)  & 0.6133(8) & 0.388466(6) & 0.993781(7) & 0.347(1)\\
      (2,4)  & 0.4651(6) & 0.460755(9) & 0.997971(2) & 0.346(2) \\
      (2,5)  & 0.4217(8) & 0.526968(8) & 0.999383(2) & 0.348(3) \\
      (2,6)  & 0.4017(4) & 0.583207(5) & 0.999797(8) & 0.346(2) \\
      (3,3)  & 0.3980(1) & 0.472670(3) & 0.997277(9) & 0.338(2) \\
      (3,4)  & 0.3715(9) & 0.548710(9) & 0.999181(1) & 0.341(1) \\
      (3,5)  & 0.3625(0) & 0.619447(1) & 0.999802(7) & 0.343(1) \\
      (3,6)  & 0.3582(8) & 0.678423(6) & 0.999952(4) & 0.344(1) \\
      (4,4)  & 0.3517(1) & 0.635157(1) & 0.999720(1) & 0.3438(5) \\
      (4,5)  & 0.3496(5) & 0.710952(4) & 0.999948(9) & 0.3455(3) \\
      (4,6)  & 0.3488(4) & 0.776258(6) &  0.999993(5)& 0.3464(9) \\
      (5,5)  & 0.3475(1) & 0.802231(3) & 0.999990(1) & 0.346(2)\\
      (5,6)  & 0.3474(7) & 0.876628(1) & 0.999999(1) & 0.347(1) \\
      (6,6)  & 0.3472(5) & 1.00000000  &  1.00000000 & 0.3472(5) \\
       \hline\hline
   \end{tabular}
   \caption{Simulation results for $p^\infty_c$, $p_j$, 
   $\theta_{\rm sites}$ and $\rho^\infty_c$ for the triangle 
   lattice.  The horizontal lines in the first column correspond 
   to cases without percolation.}
    \label{tab:pcpj}
\end{table}


\section{Conclusions}\label{sec:conclusions}

We investigate restricted-valence random sequential adsorption
in its pure and disordered versions, on the square and triangle lattices.
We show that there is no percolation for $V_{\rm max}=2$ on the square lattice. 
In other cases, Monte Carlo simulations coupled
with the Newman-Ziff algorithm are used, revealing
a continuous transition for cases $(2,3)$, $(2,4)$, $(3,3)$,
$(3,4)$,  and evidently, $(4,4)$. Finite-size scaling analysis of
$M^\prime_2, K_2$ and $Q^{js}_{ir}$ is employed to estimate the percolation
threshold. Scaling analyses show that the
critical exponents are in good agreement with the standard percolation
universality class, as one might expect given that the correlations are local. In the disordered cases, the universality class
is preserved although the percolation threshold naturally depends on the 
average valence.  

The coverage densities and jamming coverages are estimated for all
cases. On the square lattice, our result for the jamming of dimer RSA, 
$\theta = 0.906814(5)$, is in agreement with
previous estimates of Nord and Evans~\cite{No85} and 
of de Oliveira and coworkers~\cite{Ol92}, and surpasses their precision. 
The jamming fraction 
was estimated for all cases, with uncertainties on the order of $10^{-6}$. 
Our estimates for percolation thresholds in restricted-valence cases (i.e., 
some fraction of sites having $V_{max} < 4$) are $\geq 1/2$, as expected, 
except for (3,3), where we find $p_c = 0.496(2)$. For the triangle
lattice, our estimate for $p_c$ is again in agreement with the known exact 
value. The largest estimate for $p_c$ is for $(1,4)$, with 
$p^\infty_c=0.8582(5)$.
Our study also provides an estimate for the jamming coverage of dimer RSA
on the triangle lattice: $\theta = 0.914028(2)$.  This is consistent with, 
and considerable more precise than, the previous estimate of 0.9142(12) obtained 
by Perino et al.~\cite{Pe17}.
Our results for the triangle lattice are again consistent with the standard 
percolation universality class. 

We expect our results to be of use in interpreting experiments on 
irreversible deposition on substrates of reduced functionality, and/or with 
disorder, and to the question of percolation of the deposited structure.
Our study of percolation in mixed cases suggests several avenues for future work.
It should be possible, if perhaps challenging, to demonstrate mathematically that
on the square lattice, the (1,4) mixture does not percolate whereas the (2,3) does.
More generally, identifying the percolation threshold {\it surface} on the simplex
$f_1 + f_2 + f_3 + f_4 = 1$ (where the $f_i$ denote site fractions with maximum 
valence $i$ in a random mixture) is an outstanding challenge.  Similar open questions 
exist for the triangle lattice.

\section{Acknowledgments}
AF, DCS and RD and acknowledge financial support from CNPq, CAPES and 
Fapemig, Brazil. RD acknowledges support from CNPq under project 
number 303766/2016-6.

\section{Appendix: No percolation for \texorpdfstring{$V_{\rm max}=2$}{Lg}}

Consider a final configuration of dimer deposition on $\mathbb{Z}^2$ with
$V_{\rm max}=2$.  Let $C$ be the set of sites connected to the origin, $\OO$.
$C$ is the union of sites that are 1, 2, ..., $n$ steps from $\OO$, along
the path of occupied edges linking the sites to $\OO$.  Let
${\sf X}_j \in \{0,1,2\}$ be the number of sites exactly $j$ steps from $\OO$ in $C$.
(In case $C$ be a closed loop, we take $j$ as the minimum number of steps to $\OO$.)
We note that ${\sf X}_j$ is a nonincreasing sequence, with ${\sf X}_j = 0$
for $j>n$.  Let $P(n)$ be the probability, over the space of all final
configurations, that the maximum number of steps from $\OO$ in the cluster
containing the origin is $n$. We aim to show that $\exists \, c<1$ such that
$P(n) < c^n$, so that $\mathbb{E} \,n < \infty$, i.e., there is no percolation.

\begin{figure}[htbp]
  \centering
  \includegraphics[scale=0.5]{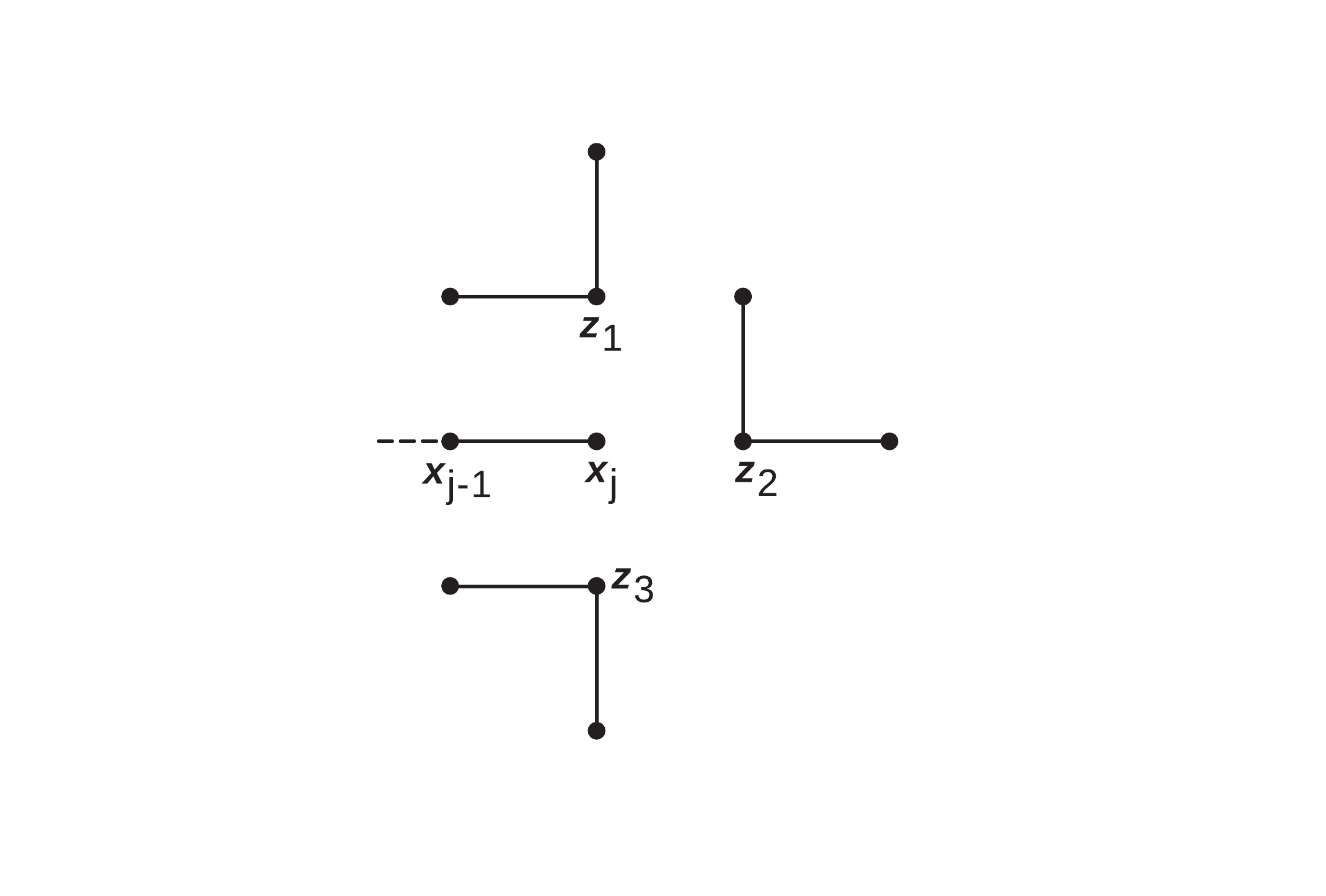}
\vspace{-2em}
  \caption{\footnotesize A configuation corresponding to event $B_j$.}
  \label{sk1}
\end{figure}

Let $A_j$ be the event ${\sf X}_j=2$.  We show that $P(A_{j+1}|A_j) < 1$.
Given $A_j$, let $x_j$ be one of the sites that are $j$ steps from $\OO$, let 
{\bf z}$_1$, {\bf z}$_2$, and {\bf z}$_3$ 
be the sites neighboring {\bf x}$_j$ distinct from {\bf x}$_{j-1}$, and let $y_i$
be the edge linking sites {\bf x}$_j$ and {\bf z}$_i$. Let $B_j$ be the event that
edges $y_1$, $y_2$, and $y_3$ are all unoccupied in the final configuration.
Occurrence of $B_j$ implies that ${\sf X}_{j+1} < 2$.  Thus, $P(B_j) > 0$ implies
$P(A_{j+1}|A_j) < 1$.

We now argue that $P(B_j) > 0$ for any $j > 0$.  A sufficient condition for
occurrence of $B_j$ is that sites {\bf z}$_1$, {\bf z}$_2$, and {\bf z}$_3$ are all
of valence two, via edges $y_m \notin  \{y_1, y_2, y_3\}$.  An example of such a
configuration is shown in Fig.~\ref{sk1}. Recall that each edge $y$ is assigned
a real time $t_y \in (0,1]$.  Initially all edges are empty, and edge $y$ becomes occupied at
time $t_y$ if permitted by the maximum-valence rules; otherwise it remains empty
for all time.

Consider a site {\bf s} and let $u_1,...,u_4$ be the edges incident on {\bf s}, with
$t_1,...,t_4$ their associated times.  Call $e({\bf s})$ the pair of edges $u_i$
having the smallest and second smallest times.  For an arbitrary edge $y$, let
{\bf s}$_1$ and {\bf s}$_2$ be its terminal sites.  A sufficient (but not necessary)
condition for $y$ to be occupied in the final configuration is that $y$ belong to
$e({\bf s}_1)$ and to $e({\bf s}_2)$.  Using the fact that the $t_j$ are i.i.d.
uniformly distributed on (0,1], one readily verifies that
Prob[$y \in e({\bf s}_1) \cap e({\bf s}_2)$] = 13/35.

\begin{figure}[htbp]
  \centering
  \includegraphics[scale=0.5]{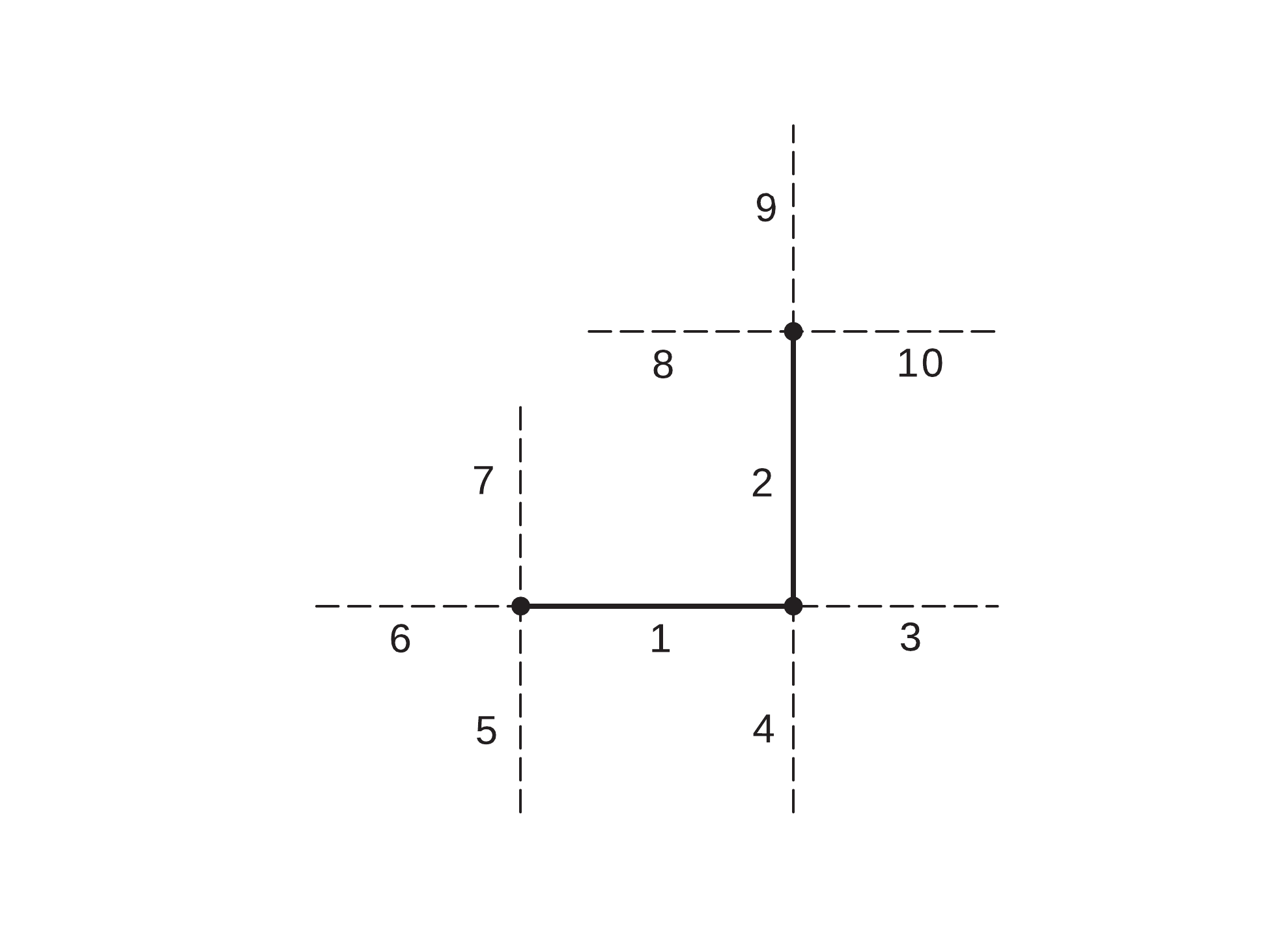}
\vspace{-2em}

\caption{\footnotesize Part of a configuration corresponding to event $B_j$.
  Numbers label the edges and associated times.
}
  \label{sk2}
\end{figure}

Now consider the local configuration shown in Fig.~\ref{sk2}.  A sufficient condition
for the edges indicated by solid lines to be occupied is that they belong to the sets
$e$ associated with the three sites upon which these edges are incident; these
conditions naturally imply certain restrictions on the $t_i$.  Integrating over the
$t_i$, considering the diverse orderings allowed by the above-mentioned constraints,
we find the probability of the two-edge configuration shown in Fig.~\ref{sk2} to be
at least $87/900 \equiv b^{1/3}$.  It follows that $P(B_j) > b > 0.$  By the same
argument, if $X_j = 1$, there is a nonzero probability that $X_{j+1}=0$. Thus there
is a finite constant $C$ such that Prob$[X_n > 0] < C(1-b)^n$ implying that the
percolation probability is zero.
Although the argument is for the square lattice, it can be adapted
via straightforward modification to arbitrary lattices of finite degree.

\vspace{-1em}

\vspace{1em}

\bibliography{bibliography.bib}

\end{document}